\documentclass{article}
\usepackage[english]{babel}
\usepackage[utf8]{inputenc}
\usepackage{amsmath}
\usepackage{amsfonts}
\usepackage{amssymb}
\usepackage{graphicx}
\usepackage{subcaption}
\usepackage{fancyhdr}
\usepackage{tabularx}
\usepackage{geometry}
\usepackage{setspace}
\usepackage[right]{eurosym}
\usepackage[printonlyused]{acronym}
\usepackage{floatflt}
\usepackage[usenames,dvipsnames]{color}
\usepackage{colortbl}
\usepackage{paralist}
\usepackage{array}
\usepackage{titlesec}
\usepackage{parskip}
\usepackage[right]{eurosym}
\usepackage{longtable}
\usepackage[colorlinks=true,linkcolor=black,urlcolor=blue,citecolor=black,pdfpagelabels=true,bookmarks=false]{hyperref}
\usepackage{mathtools}
\usepackage{booktabs,caption}
\usepackage[backend=biber,style=authortitle,citestyle=authoryear,maxbibnames=99]{biblatex}
\usepackage[flushleft]{threeparttable}
\usepackage{dsfont}
\usepackage{enumitem}
\usepackage[toc,page]{appendix}
\usepackage{rotating}

\usepackage{listings}
\makeatletter
\def\l@lstlisting#1#2{\@dottedtocline{1}{0em}{1em}{\hspace{1,5em} Lst. #1}{#2}}
\makeatother

\newcommand\posscite[1]{\citeauthor{#1}'s (\citeyear{#1})}

\numberwithin{equation}{section}
\begin{document}

\thispagestyle{empty}
\begin{center}
	\vspace*{2cm}
	\Huge
	\textbf{The Finite Sample Performance of Treatment Effects Estimators based on the Lasso}
	
	\vspace*{1cm}
	\Huge
	
	\Large
	
	\vspace*{5.5cm}
	\begin{large}
	Michael Zimmert\\
	michael.zimmert@unisg.ch\\
	Tel. +41 (0)71 224  2328\\
	
	Swiss Institute for Empirical Economic Research\\
	Universität St.Gallen | Varnbüelstrasse 14 | CH-9000 St.Gallen

	\end{large}

\end{center}

\vspace*{6cm}
	\begin{small}
	\end{small}
\pagebreak

\setcounter{page}{1}
\onehalfspacing
\allowdisplaybreaks
\begin{abstract}
\noindent
This paper contributes to the literature on treatment effects estimation with machine learning inspired methods by studying the performance of different estimators based on the Lasso. Building on recent work in the field of high-dimensional statistics, we use the semiparametric efficient score estimation structure to compare different estimators. Alternative weighting schemes are considered and their suitability for the incorporation of machine learning estimators is assessed using theoretical arguments and various Monte Carlo experiments. Additionally we propose an own estimator based on doubly robust Kernel matching that is argued to be more robust to nuisance parameter misspecification. In the simulation study we verify theory based intuition and find good finite sample properties of alternative weighting scheme estimators like the one we propose.
\end{abstract}
\section{Introduction}
In observational studies treatment effects estimation is a delicate task. Unlike experimental designs such settings heavily rely on what is called a conditional independence or exogeneity assumption. Credible identification of causal effects requires that the biasing effect of confounding variables is purged out by controlling for them. For example, evaluating active labour market programs typically aims at estimating the causal effect of a certain policy like a training program for unemployed on employment outcomes. However, in situations that were not explicitly designed for scientific evaluation of a certain policy the major challenge arises from the fact that policy assignment can in general not be regarded as random. In our example unemployed that are more educated might self select into training programs if participation is voluntary and not controlling for potential confounders can heavily bias the effect one wants to measure. Thus, the selection of the `right' covariates is at the heart of observational studies econometrics.\\
Recent advances in the causal inference literature cope with problems where the dimension of the covariate space is large -- potentially much larger than the number of observations (\textcite{Belloni_Chernozhukov_Hansen_2014}, \textcite{Farrell_2015}, \textcite{Kennedy_2016}, \textcite{Athey_Imbens_Wager_2018}, \textcite{Chernozhukov_Chetverikov_Demirer_Duflo_Hansen_Newey_2017}). This issue may arise in two situations that should be distinguished. First of all, administrative and other datasets make a huge amount of potential confounders available to researchers. Second, including covariates not only in levels but also as higher order polynomials and interactions might help to enhance credible causal identification in observational studies. While the first situation might be seen as a symptom of the age of `Big Data', the second situation basically represents the well-known case of global nonparametric approximation. Being capable to cope with the latter situation without being exposed to the `curse of dimensionality' would be desirable since it would ease functional form dependence in causal inference. In both cases this brings up the challenge that in settings where the dimension of the covariate space is large compared to the number of observations, the researcher has to select the covariates that should enter the model. Typically, applied researchers select their empirical model based on some theoretical prediction, data availability, or in an ad-hoc fashion just having some feeling that a variable should be included. Estimation methods like Lasso labelled as supervised `machine learning' are able to reduce the dimension of the model and find the variables that predict another variable best. Clearly, introducing a machine learning step brings up new issues when thinking about inference. At a first glance, it might look like a rather philosophical question if the uncertainty of an estimator is influenced by the decision of the researcher which model to choose. Actually the potential additional uncertainty coming from model selection can only be correctly estimated if the selection step is performed explicitly -- that is within the controllable statistical framework of machine learning -- and not ad-hoc.

Formally the problem under investigation extends \posscite{Rubin_1974} causal framework. The often cited `fundamental problem of causal inference' (\textcite{Holland_1986}) is that in the case of binary treatments one observes only one of the two potential realizations. An unemployed either decides to participate in a training program or not. We never observe the counterfactual situation. Let $D_i\in\{0,1\}$ denote the treatment status of observation $i=1,...,n$ and let $Y_i$ denote the observed outcome of interest for that observation. Then the observed outcome can be written as a function of potential outcomes
\begin{align}
Y_i=\begin{cases}
Y_i(0) & \mbox{if} \, D_i=0\\
Y_i(1) & \mbox{if} \, D_i=1
\end{cases}. 
\end{align}
In an observational study the triple $(Y_i,D_i,X_i)$ is observed where $X_i$ is a vector of potential confounders. The setting of interest involves an untestable conditional independence assumption
\begin{align}\label{eq:CIA}
Y_i(0),Y_i(1)\perp D_i | X_i \quad \text{(CIA)}.
\end{align}
Thus, knowing the confounder space $X$ solves the selection into treatment problem discussed above. Statistics of interest in typical applications are given by 
\begin{align}
\text{ATE}&=E[Y_i(1)-Y_i(0)] \quad \text{and}\\
\text{ATET}&=E[Y_i(1)-Y_i(0)|D_i=1].
\end{align}
Only under the necessary condition that the CIA holds, the difference in the potential outcomes for the whole population (ATE) or the population that has received treatment (ATET) can be identified and the measured statistics may be interpreted as reflecting causal effects. Several estimator classes ranging from the classical parametric outcome regression to inverse probability weighting (IPW) (\textcite{Hirano_Imbens_Ridder_2003}), K-nearest-neighbour matching (\textcite{Abadie_Imbens_2006}), propensity score matching (\textcite{Heckman_Ichimura_Todd_1998}, \textcite{Dehejia_Wahba_2002}) and optimal weighting approaches (notably \textcite{Graham_Campos_Egel_2012}, \textcite{Hainmueller_2012} and \textcite{Zubizarreta_2015}) have been proposed in the literature. While the traditional outcome regression potentially suffers from functional form misspecification, the other approaches balance the outcome distributions in the treatment and control group conditional on the covariates. In practice the latter approaches enable semiparametric estimation of treatment effects. A standard strategy is to estimate weights parametrically and then identify the treatment effect of interest nonparametrically using the reweighted outcomes. For example \textcite{Rosenbaum_Rubin_1983} show that conditioning on the so called propensity score defined as $e(x_i)=Pr(D_i=1|X_i=x_i)$ is as good as conditioning on the covariates directly. \textcite{Angrist_Pischke_2009} among others find that indeed all econometric estimators can be regarded as reweighting observed outcomes with weights being a function of the covariates included. Not relying on a functional form in the nonparametric step is more honest in the sense that it prevents the researcher to draw conclusions from extrapolation in areas where she just does not have any information on. In other words, overlap of the covariate distributions is required which can be expressed as 
\begin{align}
0<e(x_i)<1 \quad \forall \,i \quad \text{(overlap)}.
\end{align}
For good surveys on the literature with low-dimensional covariate spaces see \textcite{Imbens_Wooldridge_2009} and \textcite{Athey_Imbens_2017}.\\
In the context of high-dimensional data the additional challenge arises that $X\in\mathbb{R}^p$ and $p>>n$. Thus, we allow for a case that was implicitly ruled out for standard parametric or nonparametric estimation techniques. The pioneering work of \textcite{Belloni_Chernozhukov_Hansen_2014}, \textcite{Farrell_2015} and \textcite{Belloni_Chernozhukov_FernandezVal_Hansen_2017} uses the semiparametric efficient influence function theory behind the original doubly robust estimation techniques of \textcite{Robins_Rotnitzky_1995} and \textcite{Hahn_1998} for the integration of machine learning methods to the framework of treatment effects identification. Since both the outcome and the treatment equation are predicted using machine learning (for an overview of the different algorithms see \textcite{Hastie_Tibshirani_Friedman_2009}), they call their estimation technique double machine learning. De facto they use the classical IPW estimator that is augmented with some outcome regression terms (AIPW). \textcite{Athey_Imbens_Wager_2018} develop what they call approximate residual balancing (ARB) under a different regime of assumptions. In contrast to AIPW their estimator does not rely on propensity score estimation, but uses suggestions from the optimal weighting literature to weight the debiasing residual from the outcome equation term.

The contribution of this article is twofold. In a first step we outline the statistical concepts necessary to compare different reweighting schemes in the context of high-dimensional covariate spaces. Particularly, we review the new literature on high-dimensional treatment effects estimation from the perspective of traditional semiparametric theory. Understanding the latter as the basis for the machine learning based estimators enables to link them with popular methods like matching. Guided by asymptotical considerations we focus on the efficient influence function approach to treatment effects estimation. This motivates the discussion of recent contributions of \textcite{Belloni_Chernozhukov_Hansen_2014} and \textcite{Athey_Imbens_Wager_2018}. Further we also present a modified version of the radius matching estimator of \textcite{Lechner_Miquel_Wunsch_2011} that is claimed to have good statistical properties while being able to address some finite sample concerns of the efficient influence function approach. However, unlike ARB it is suitable for non-linear machine learning estimators. The second step is to validate our basic claims via various simulation experiments. We show that alternative reweighting schemes like the one we propose perform well in finite sample.

The rest of the article is organized as follows. In the two following sections we will motivate a semiparametric estimation structure that is suitable for the incorporation of machine learning estimators. Section \ref{sec:naive} studies a naive approach to combine treatment effects estimation with Lasso prediction. The failure of this approach will motivate the so called doubly robust methods discussed in section \ref{sec:doublyrobust}. We will theoretically review the different features and assumptions behind AIPW, ARB and our matching based estimator. A second part in section \ref{sec:montecarlo} then uses various Monte Carlo experiments to investigate the properties of the different estimators in finite sample. The last section concludes.
\section{The Problem: Single Equation Approaches with the Lasso}\label{sec:naive}
\begin{figure}[h!]
\caption{Empirical distribution of IPW with predicted propensity score}
\label{fig:ipw_fail}
\begin{subfigure}{.5\textwidth}
\includegraphics[scale=0.5]{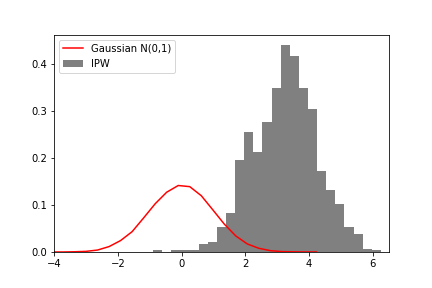}
\caption{$R^2_D=0.5$, $R^2_Y=0.2$}
\end{subfigure}
\begin{subfigure}{.5\textwidth}
\includegraphics[scale=0.5]{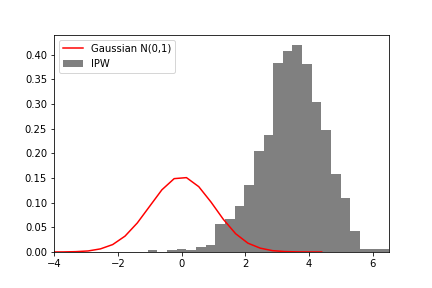}
\caption{$R^2_D=0.5$, $R^2_Y=0.8$}
\end{subfigure}
\end{figure}

Given previous considerations at a first glance it seems to be a good idea to apply a machine learner directly to a standard treatment effects estimator. To make this point a bit more concrete, assume that a normalized IPW estimator (see \textcite{Busso_DiNardo_McCrary_2014} for a discussion) is used to estimate
\begin{align}
\hat{\text{ATE}}&=\frac{1}{n_1}\sum_{i=1}^nD_i\hat{W}_i^{IPW}Y_i-\frac{1}{n_0}\sum_{i=1}^n(1-D_i)\hat{W}_i^{IPW}Y_i \quad \text{where} \label{eq:IPW}\\
\hat{W}_i^{IPW}&=\mathds{1}\{D_i=1\}\frac{1}{\hat{e}(X_i)}\Bigg/\sum_{i=1}^n\frac{1}{\hat{e}(X_i)}+\mathds{1}\{D_i=0\}\frac{1}{1-\hat{e}(X_i)}\Bigg/\sum_{i=1}^n\frac{1}{1-\hat{e}(X_i)} \label{eq:IPWweights}
\end{align}
and $n_1$ and $n_0$ are the number of treated and control observations. In the high-dimensional case when $p>>n$ standard methods to estimate the propensity score will be infeasible. However, it can in principle be estimated using any supervised machine learning technique. Lasso always optimizes a $L_1$-regularized version of the standard loss function. For the case of logistic regression the problem can be written as
\begin{align}
\hat{\beta}=\text{arg}\,\text{min}_{\beta}\left(\sum_{i=1}^n(\log(1+e^{X_i\beta})-D_iX_i\beta)+\lambda\sum_{j=1}^p|\beta_j|\right).
\end{align}
When the tuning parameter $\lambda$ is zero, the problem transforms to the standard logistic regression optimization. In settings where $p\rightarrow n$ the in-sample fit of the unpenalized model can become arbitrary large by just increasing the number of covariates. Thus, for any $\lambda>0$ the goal is not to optimize the in-sample fit but to minimize the out-of-sample mean squared error (MSE). Thus, Lasso instead of fitting the model optimally in sample aims at achieving good out-of-sample predictions by avoiding overfitting. This means that including additional covariates in the model is penalized in the minimization problem. In fact, the $L_1$-norm will shrink some of the coefficients to zero. It follows that with Lasso one achieves the dimensionality reduction necessary to fit a propensity score model in the case when $p>>n$. When a high $p$ is achieved by series expansion Lasso \textit{approximates} global nonparametric methods. In some sense machine learning methods overcome the curse of dimensionality introduced by nonparametric methods. However, it is important to notice that these methods should not be confused with nonparametric approaches. Rather by shrinking some parameters to zero Lasso might be seen as a data-driven compromise between fully flexible nonparametric and rich parametric methods. In principle, it is therefore very appealing to fit the propensity score model in a regularized form since this approach represents a trade off between functional form dependence and the dimensionality of the problem.\\
It follows from this discussion that the choice of the tuning parameter is crucial for the performance of the Lasso. Theoretical results are typically derived under a sparsity assumption which means that only few covariates $s$ really matter in the model. Additionally there are usually some assumptions that guarantee that the correlation between feutures is not too strong. For example \textcite{Bickel_Ritov_Tsybakov_2009} and \textcite{Belloni_Chernozhukov_2011} show that for $p>n$ under such a set of assumptions the empirical $L_2$ prediction norm\footnote{In the linear model implied by the Lasso the prediction norm upper bounds the prediction error. This result follows from the triangle inequality and suggests the same convergence rates for the prediction error. For a discussion see \textcite{Belloni_Chernozhukov_2011}.} is bounded by
\begin{align}\label{eq:lassoconvergence1}
\left\Vert\hat{\beta}-\beta_0\right\Vert_{2,n}=\mathcal{O}\left(\sqrt{\frac{s\log p}{n}}\right).
\end{align}
Despite the asymptotic appeal of such well-defined convergence rates, practitioners typically use $k$-fold cross-validation as it mimics the idea of optimizing out-of-sample MSE. In a recent paper \textcite{Chetverikov_Liao_Chernozhukov_2017} show that under Gaussian errors the prediction norm is bounded by
\begin{align}\label{eq:lassoconvergence2}
\left\Vert\hat{\beta}-\beta_0\right\Vert_{2,n}=\mathcal{O}\left(\sqrt{\frac{s\log p}{n}}\log^\frac{7}{8}(pn)\right).
\end{align}
The authors also show that by relaxing the normality assumption lower convergence rates may be obtained. It follows that asymptotically the price to pay for not relying on theory based tuning parameters are worse convergence rates for the Lasso, though we acknowledge that in finite samples the more data-driven cross-validation based approaches might be less sensitive against violations of the main Lasso assumptions. Alternatively, practitioners also choose the tuning parameter by taking a value that is one standard error to the right of the minimal cross-validation criterion and thus select a smaller than the cross-validation optimal model. Originally the method was proposed for regression trees by \textcite[chapter 3]{Breiman_Friedman_Olshen_Stone_1984} without giving any theoretical justification. \textcite[chapter 7]{Buehlmann_vanderGeer_2011} show that the true model is nested within the chosen model if the tuning parameter is determined via cross-validation. This result immediately implies that when having variable selection and not prediction in mind a larger penalty term might be more appropriate. These issues are discussed in more detail in section \ref{sec:montecarlo} of this analysis.\\
Despite the very appealing nature of the Lasso, we will develop two arguments why a procedure that uses the machine learning prediction of the propensity score solely in a second stage is not a valid estimation strategy.\\
\begin{enumerate}
\item \textbf{The econometric argument}: Consistency of a treatment effects estimator implicitly requires that variables which jointly have an effect on the treatment and the outcome should be included in the model. These variables are often called confounders. If variable selection would be perfect, the active set -- in our case all confounders that have a `true' effect on the propensity score -- should be included in the model. Clearly, this would also include those variables that also have an effect on the outcome. \textcite[chapter 7]{Buehlmann_vanderGeer_2011} indeed show that Lasso is consistent for model \textit{selection} under quite restrictive assumptions. Probably the one that is most unlikely in real applications is the so called `beta-min' condition that basically states that the active set has coefficients that are (in absolute value) larger than a certain lower bound. More precisely the `beta-min' condition requires that variables which truly belong to the model have coefficients that are not too small. This
guarantees that the Lasso can detect all relevant variables. The second necessary condition concerns the covariance structure of the variables potentially to be selected and is called the neighbourhood stability or irrepresentable condition. In particular, the covariance structure of the active variables to the nonactive variables should be bounded between zero and one relative to the covariance among the active variables. One could interpret this as the relative dependence between `true' and unnecessary variables to be relatively small. Intuitively, if two variables in the active and the nonactive set are highly correlated then it will be hard for the Lasso to select the `true' variable from the active set since it can be well explained by the irrelevant variable from the nonactive set. In practice these conditions are likely to fail. The failure of model selection consistency in the presence of small but relevant confounders when using only the outcome equation is particularly stressed by \textcite{Belloni_Chernozhukov_Hansen_2014}. Confounders that have a negligible effect on the outcome but determine the treatment are hardly detected when using the outcome equation only. The authors suggest double selection that is to perform model selection on both the outcome and the treatment equation and then take the union of selected covariates to be included in the second stage estimation step.
\item \textbf{The statistical argument}: More generally \textcite{Leeb_Poetscher_2005} comment on inference after model selection methods. They find that the first stage model selection step has some influence on the inference of the estimator of interest. This is often overlooked since there is the result that under consistent model selection asymptotically the same results on inference apply whether there was model selection in a previous step or not (\textcite{Poetscher_1991}). Indeed, this is closely linked to what is discussed in the literature as the oracle property which was proposed as a quality indicator by \textcite{Fan_Li_2001}. \textcite{Leeb_Poetscher_2005} confirm the validity of this statistical analysis but doubt the usefulness of it. Their main point is that these results are valid pointwise but not uniformly. In statistics normally results that are shown to be valid asymptotically can be transformed to some finite sample application, and results achieved there can then be assumed to converge to their asymptotic counterparts if the sample size increases. However, strictly speaking, this requires uniform convergence whereas proofs are often performed in a pointwise convergence setting. Thus, a proof that is valid under pointwise convergence may say something about asymptotics but not about the behaviour of the estimator in finite sample. The asymptotic properties might just have no finite sample counterparts. The often cited example in this context is Hodges' estimator introduced by \textcite{LeCam_1953} for pedagogical reasons. In its most general version it can be written as
\begin{equation}\label{eq:hodge1}
\breve{\theta}_n=\begin{cases}
(1-\alpha)c+\alpha\hat{\theta}_n & \mbox{if}\, \left|\hat{\theta}_n-c\right|\leq a_n \\
\hat{\theta}_n & \mbox{if}\, \left|\hat{\theta}_n-c\right|> a_n \\
\end{cases}
\end{equation}
where $\alpha\in[0,1]$, $c$ is a constant and $\hat{\theta}_n$ is an asymptotically linear, efficient and consistent estimator. In \posscite{LeCam_1953} original specification he shows the remarkable result that from a pointwise analysis $\breve{\theta}_n$ is also consistent but achieves a smaller asymptotic variance than the original estimator. How is this related to inference after model selection? Following \textcite{Wu_Zhou_2016} assume that $\alpha=0$. Then Hodges' estimator (\ref{eq:hodge1}) looks like a single hypothesis test estimation procedure. Either a model is accepted and then estimated with the estimator $\hat{\theta}$ or it is rejected and all coefficients are assumed to be zero. The two cases can be achieved with the Lasso by either choosing $\lambda$ as being equal to zero or by setting the tuning parameter sufficiently large such that all coefficients are shrunken to zero and only a constant term $c$ remains. Observing this leads \textcite{Leeb_Poetscher_2008} to the claim that inference after model selection is nothing else than the return of Hodges' estimator with all undesired properties notably the lack of uniform convergence. A simple model selection step prior to identification will therefore lead to inconsistent estimators and invalid inference.
\end{enumerate}
Figure \ref{fig:ipw_fail} shows the distribution of the ATE for IPW with a logistic Lasso model for predicting the propensity score. The distribution was generated with a Monte Carlo experiment using a design where both the outcome and the treatment equation are determined by an approximately sparse sequence of covariates for the high-dimensional case $n=2000$ and $p=2000$ (design 1; see section \ref{sec:montecarlo} for details). The distribution is recentered at zero such that an unbiased estimator would have mean zero. Obviously, the estimator is heavily biased and probably non-normal in both specifications. If the relation between the covariates and the outcome equation gets stronger and $R^2_Y$ is increased from 0.2 to 0.8 the bias gets stronger since now covariates left out by the Lasso when estimating the propensity score potentially have a stronger effect on the outcome. Therefore, the need to include them in the model gets more urgent and the failure when not doing so gets heavier. The result complements the results in \textcite{Belloni_Chernozhukov_Hansen_2014} for the case of treatment effects estimation based on the propensity score.\\
To conclude, the naive idea of just applying machine learning tools to econometrics turns out to be a delicate approach when aiming at inference and causal interpretation. Broadly speaking, the reason for this is that for valid inference typically both the outcome and the treatment equation have to be used -- a feature so called doubly robust estimators incorporate. We proceed by reviewing their properties in the next section.
\section{The Solution: Doubly Robust Estimators}\label{sec:doublyrobust}
\subsection{Semiparametric Theory}
The beauty of the so called doubly robust treatment effects estimators is that they can be directly derived from semiparametric theory. In fact, all other classes of treatment effects estimators can then be regarded as `only' adjusting their weighting and bias-adjustment terms to some perceived finite sample needs.\\
In the following we will continue with a short review of semiparametric theory that will prove to be useful to understand causal treatment effect estimation using machine learning techniques. In particular, considering the very general question how an asymptotically optimal estimator should look requires the notion of influence functions or curves. Let $\psi$ be a regular asymptotically linear estimator for estimating the effect of $D$ on $Y$ in a parametric outcome regression model. Then its influence function $\phi$ for the random realization $Z_i=(Y_i,D_i,X_i)$ is determined by
\begin{align}
\hat{\psi}-\psi=\frac{1}{n}\sum_{i=1}^n\phi(Z_i)+o(n^{-\frac{1}{2}}).
\end{align}
Knowing the influence function of the estimator will enable the researcher to infer the asymptotic properties. Consistency implies that $E[\phi(Z)]=0$ and the asymptotic variance will then be given by $E[\phi(Z)\phi(Z)^T]$. There are well-known results for parametric estimators. For example maximum likelihood estimation (MLE) implies that the Hessian converges to the product of scores, the Cram\'{e}r-Rao lower bound is achieved (see for example \textcite[chapters 12 and 13]{Wooldridge_2010}) and the influence function can be estimated using
\begin{align}
-\left(\sum_{i=1}^ns_{\theta_0}(Z_i)s_{\theta_0}^T\right)^T\left(\sum_{i=1}^ns_{\theta_0}(Z_i)\right).
\end{align}
The problem for semiparametric treatment effects estimators, however, is slightly more sophisticated for two reasons.
\begin{enumerate}[label=(\roman*)]
\item As discussed in section \ref{sec:naive} treatment estimators reweight the observed outcomes. However, the weights for this step are obtained as some function of observed covariates and the issue arises if and how the asymptotic behaviour of treatment effects estimators will depend on this nuisance parameter estimation step.
\item Moreover, the weights depend on nuisance parameters of infinite dimension. For example in the case of IPW estimation the inverse of the propensity score is a conditional expectation and therefore a function of $X$. In the following we will see that also conditional expectations of $Y$ given $X$ will be considered as being part of the nuisance parameter space. Conditional expectations are random variables themselves and are therefore of infinite dimension.
\end{enumerate}
Conceptually, semiparametric theory builds on the separation of the two problems (i) and (ii). We proceed by first considering a problem where the nuisance parameter $\eta$ is of finite dimension in section \ref{sec:finitenuisance} and investigate the asymptotic effects of using the estimated nuisance parameter as an input in the identification step. Linking this to the analysis in \ref{sec:infinitenuisance} that translates the problem to infinite-dimensional nuisance settings requires that we formulate it very generally. Unlike for standard outcome regression problems we replace the concepts of Euclidean with Hilbert space vector geometry. Roughly speaking, in contrast to Euclidean spaces a Hilbert space $\mathcal{H}$ also captures the case when vectors that span this space have elements that are random variables itself. Most importantly the smallest distance between a Hilbert space vector $h$ and a linear subspace of the Hilbert space $\mathcal{U}$ is orthogonal to any other element of that subspace $\mathcal{U}$. The unique point in $\mathcal{U}$ that is closest to $h$ is then defined as the projection $\Pi(h|\mathcal{U})$ of $h$ onto the subspace. For a more rigorous treatment of these concepts we refer to \textcite[chapter 2]{Tsiatis_2006} and \textcite[chapter 1]{vanderLaan_Robins_2003}.
\subsubsection{Influence functions with finite dimensional nuisance}\label{sec:finitenuisance}
Suppose the estimation problem generalizes to $\psi=\{\theta^T, \eta^T\}^T$. Specifically let $\theta$ be the statistic of interest like for example the ATE and $\eta$ the first stage nuisance parameter like for example the propensity score. Since we are interested in the asymptotic behaviour of $\theta$, our goal is to derive the moments of its influence function. Like in the standard M-estimation setting, scores might be particularly helpful. Following \textcite[chapter 3]{Tsiatis_2006} define the scores of the log-likelihood with respect to the two parameters as
\begin{align}
s_{\theta}(Z,\psi_0)&=\frac{\partial \log p_z(z,\psi)}{\partial \theta}|_{\psi=\psi_0} \quad \text{and}\\
s_{\eta}(Z,\psi_0)&=\frac{\partial \log p_z(z,\psi)}{\partial \eta}|_{\psi=\psi_0}.
\end{align}
Now suppose both score functions span spaces which are denoted as tangent spaces $\mathcal{T}_\theta\subset \mathcal{H}$ and $\mathcal{T}_\eta\subset \mathcal{H}$ where the latter is denoted the nuisance tangent space. A fundamental result of semiparametric statistics is then that all influence functions reside in the orthogonal complement of the nuisance tangent space, i.e. $\phi\in\mathcal{T}_\eta^\perp$. In fact an influence function of $\theta$ is orthogonal to the score function of $\eta$ and $E[\phi(Z)s_\eta(Z,\psi_0)]=0$ (for details see \textcite[chapter 3]{Tsiatis_2006}).\\
Since the variance of the estimator is the second moment of its influence function, the influence function with the lowest second moment is of particular interest. The efficient influence function is given by the projection of any influence function on the whole tangent space $\phi_{eff}=\Pi(\phi|\mathcal{T})$. Thus, the efficient influence function incorporates all the information coming from the first order conditions of the parameters. More precisely, the efficient influence function turns out to be a function of the efficient score
\begin{align}
\phi_{eff}(Z)&=(s_{eff}s_{eff}^T)^{-1}s_{eff} \quad \text{with} \label{eq:effif} \\
s_{eff}&=s_\theta-\Pi(s_\theta|\mathcal{T}_\eta)=\Pi(s_\theta|\mathcal{T}_\eta^\perp) \label{eq:effscore}.
\end{align}
Notice that this rather abstract formulation is well-known in parametric M-estimation. In contrast to the case without a first-stage estimation step of the nuisance parameter, the asymptotically optimal estimator is now defined in terms of the efficient score. The projection captures the effect of the nuisance parameter estimation step on the score of the parameter of interest. The efficient score in this sense represents a quantity where this effect is purged out. Thus, if estimating the nuisance parameter has no effect on the score of the parameter of interest, i.e. $\Pi(s_\theta|\mathcal{T}_\eta)=0$ then the standard MLE result without nuisance parameter applies.
\subsubsection{Influence functions with infinite-dimensional nuisance parameters}\label{sec:infinitenuisance}
The results (\ref{eq:effif}) and (\ref{eq:effscore}) can now be generalized for the case where the nuisance parameter is of infinite dimension. Fortunately, \textcite{Tsiatis_2006} formulates the problem such that the results depend on projections on subspaces of the Hilbert space. Therefore, introducing nuisance parameters of infinite dimension does not require a fundamentally new formulation of the problem. Instead one can use so called parametric submodels as  methodological devices to derive the efficient influence function with infinite nuisance. In particular one can imagine a semiparametric model $\mathcal{P}$ as a set of densities such that
\begin{align}
\mathcal{P}=\{p(Z,\theta,\eta)\}.
\end{align}
A submodel of $\mathcal{P}$ with finite-dimensional nuisance such that the true data-generating density is contained in the set of these submodels is then defined as the parametric submodel. More formally let the set of possible distributions within a submodel be given by $\{P_\epsilon : \epsilon\in \mathbb{R}\}$. If $\epsilon=0$ the parametric submodel equals the true distribution of the semiparametric model (\textcite{Kennedy_2016}). The parametric submodel is a methodological trick that allows to approximate the semiparametric model arbitrarily close but eases the mathematical concepts required drastically. In particular, an estimator for $\theta$ is a valid estimator for a semiparametric model if it is also a valid estimator for every parametric submodel of that semiparametric model. If every $\hat{\theta}$ in a semiparametric model is contained in the set of semiparametric submodels, this also has to apply for every influence function of every estimator for $\theta$. It follows that every influence function of $\theta$ is contained in the set of influence functions of the parametric submodels. This reasoning also carries over to efficiency considerations. If every influence function of the semiparametric model is contained in the influence functions of the parametric submodels, the minimal variance of a semiparametric model that can be achieved is the supremum over all parametric efficient influence function variances.\\
Efficient influence functions can again be derived by constructing the nuisance tangent space. For that purpose one can first of all define score functions for the parametric submodels that span tangent spaces. The closure of these tangent spaces is then the tangent space of the semiparametric model. Thus, we obtain $\mathcal{T}_\eta$ again and the efficient influence function can again be derived by using its efficient score representation.
\subsubsection{An Efficient Influence Function for semiparametric treatment effects estimation}
We will now make this last point a bit more concrete by showing how it can be used to check if a proposed influence function is an efficient influence function. A fundamental result that bridges the gap between the rather abstract concepts discussed above and concrete estimation techniques is given by \textcite{Newey_1990}, \textcite{Bickel_Klaassen_Ritov_Wellner_1998} and \textcite{Newey_1994}. They show that every influence function of a regular and asymptotically linear estimator has to obey
\begin{align}
\frac{\partial \theta(P_\epsilon)}{\partial \epsilon}|_{\epsilon=0}=E[\phi s_\epsilon(Z)].
\end{align} 
Thus, one needs to know the pathwise derivative of the estimator of interest and the score of the parametric submodel to check if a proposed function is indeed an influence function of the parameter of interest. The efficient influence function for the binary semiparametric ATE estimation problem is given by
\begin{align}\label{eq:effifate}
\begin{split}
\phi_{eff}=&\frac{D(Y-E[Y|D=1,X])}{e(X)}-\frac{(1-D)(Y-E[Y|D=0,X])}{1-e(X)}+E[Y|D=1,X]\\
\\
&-E[Y|D=0,X]-E[Y(1)-Y(0)].
\end{split}
\end{align}
\textcite[proof of theorem 1]{Hahn_1998} indeed shows that equation (\ref{eq:effifate}) is an influence function for the binary treatment effects estimation problem (see also \textcite{Kennedy_2016}). Since (\ref{eq:effifate}) is an element of the tangent space, results from section \ref{sec:finitenuisance} suggest that the proposed form is also the efficient influence function.\\
Obviously, the approach presented here is unsatisfactory in the sense that the form of the efficient influence function has to be guessed and can then be checked to obey the properties of an efficient influence function. The theory for average treatment effects estimation is well-established. However, in other contexts developing estimators directly from asymptotic theory might be especially appealing. For the case of treatment effects estimation \textcite[chapter 13]{Tsiatis_2006} uses the fact that influence functions for a known nuisance are much easier to derive. He develops a theoretical framework how these full-data results can be transformed to yield results when the nuisance has to be estimated.\\
Note that due to the linear form of (\ref{eq:effifate}), the efficient influence function for the ATE equals its efficient score. Recently, \textcite{Chernozhukov_Chetverikov_Demirer_Duflo_Hansen_Newey_2017} use the work of \textcite{Neyman_1979} to obtain orthogonal scores for various estimation problems. In different frameworks \ref{eq:effscore} can be derived by adjusting the original score with the projection term. For the case of ATE estimation this is equivalent to adjusting the original score as suggested by \textcite{Newey_1994}.
\subsection{Doubly Robust Estimation and Machine Learning}
A natural estimator using the moment condition $E[\phi_{eff}]=0$ implied by (\ref{eq:effifate}) is
\begin{align}
\hat{\text{ATE}}=&\frac{1}{n}\sum_{i=1}^n\bigg[\frac{D_iY_i}{\hat{e}(X_i)}-\frac{(1-D_i)Y_i}{1-\hat{e}(X_i)}-\frac{D_i\hat{E}[Y_i|D_i=1,X_i]}{\hat{e}(X_i)}+\frac{(1-D_i)\hat{E}[Y_i|D_i=0,X_i]}{1-\hat{e}(X_i)} \label{eq:aipw}\\
&+\hat{E}[Y_i|D_i=1,X_i]-\hat{E}[Y_i|D_i=0,X_i]\bigg] \nonumber \\
=&\frac{1}{n}\sum_{i=1}^n\hat{\text{E}}[Y_i|D_i=1,X_i]-\frac{1}{n}\sum_{i=1}^n\hat{\text{E}}[Y_i|D_i=0,X_i]+\frac{1}{n_1}\sum_{i=1}^nD_i\hat{W}_i^{IPW}(Y_i-\hat{\text{E}}[Y_i|D_i=1,X_i]) \nonumber \\
&-\frac{1}{n_0}\sum_{i=1}^n(1-D_i)\hat{W}_i^{IPW}(Y_i-\hat{\text{E}}[Y_i|D_i=0,X_i]) \nonumber
\end{align}
where we advice to use standardized weights (see \textcite{Busso_DiNardo_McCrary_2014}) as in (\ref{eq:IPWweights}). Originally termed as doubly robust estimation techniques this class of estimator goes back to papers by \textcite{Robins_Rotnitzky_Zhao_1994} and \textcite{Robins_Rotnitzky_1995}. The first terms are the same as for the IPW estimator while the latter terms are weighted representations of the outcome regressions for the treated and the nontreated sample. Therefore, (\ref{eq:aipw}) is also labelled as Augmented IPW (AIPW). Rearranging the estimator as in appendix \ref{app:aipw} reveals that the estimator incorporates terms like
\begin{align}\label{eq:aipwmom}
\frac{\left(E[D|X]-e(X)\right)\left(E[Y(1)|X]-E[Y|D=1,X]\right)}{e(X)}.
\end{align}
In order to achieve consistency, they have to converge to zero in expectation. Thus, either the propensity score model or the outcome regression model or both have to be correctly specified to achieve consistency. It follows that the estimator is protected against misspecification if at least one model is correct. This feature lead \textcite{Scharfstein_Rotnitzky_Robins_1999} to call such an estimator doubly robust.\\
Particularly, double robustness is a result that stems from the fact of combining two moment conditions multiplicatively. Actually this is exactly what follows from the definition of the efficient influence function as the projection of any influence function on the whole tangent space. Incorporating all information in the case of treatment effects estimation means to model the selection into treatment as well as the outcome regression step as nuisance parameters.\\
In the context of machine learning estimators the feature of efficient scores relying on these double moment conditions is particularly helpful. In section \ref{sec:naive} the failure of combining IPW estimation with machine learning estimation of the propensity score was described. In contrast to the efficient influence function motivated AIPW that takes into account all nuisances, IPW solely relies on the propensity score model. The corresponding terms to (\ref{eq:aipwmom}) are given by
\begin{align}\label{eq:ipwmom}
\frac{(E[D|X]-e(X))E[Y(1)|X]}{e(X)}.
\end{align}
Now using the Lasso convergence rates as in (\ref{eq:lassoconvergence1}) or (\ref{eq:lassoconvergence2}) indicates that such terms will not exhibit $\sqrt{n}$-convergence. In particular, (\ref{eq:ipwmom}) diverges if it is scaled up due to the fact that $s\log p\rightarrow\infty$. Hence, it is the decreased Lasso convergence rate that leads to the failure of IPW in this setting. On a more profound level this explains the bad behaviour of IPW with Lasso predicted propensity scores as shown in figure \ref{fig:ipw_fail}. \textcite{Belloni_Chernozhukov_Hansen_2014} therefore propose to use AIPW where the multiplicative structure of the moment term (\ref{eq:aipwmom}) protects against the divergent behaviour of a single Lasso prediction. In particular, using the Lasso convergence rate (\ref{eq:lassoconvergence1}) from the plug-in tuning parameters and rescaling term (\ref{eq:aipwmom}) results in a term of order $\frac{s\log p}{\sqrt{n}}$. For consistency this term has to vanish which implies a sparsity condition $\frac{s^2\log^2p}{n}\rightarrow 0$. Similarly using the cross-validation convergence rate (\ref{eq:lassoconvergence2}) implies the more restrictive sparsity condition $\frac{s^2\log^2 p\log^\frac{7}{2}(pn)}{n}\rightarrow 0$. Investigating the properties of their estimator more deeply, \textcite{Chernozhukov_Chetverikov_Demirer_Duflo_Hansen_Newey_2017} propose different ways to incorporate sample splitting in the estimation approach. Instead of using the same sample for prediction and second stage estimation, they suggest to partition the sample in $k$ different subsamples such that the first stage nuisance prediction step is performed on $k-1$ subsamples and the less sophisticated second stage ATE estimation step on the remaining subsample. Iterating over the different left out possibilities and taking the average of all $k$ estimators finally results in reduced sparsity conditions.\\
A potential problem with AIPW, independent of how the nuisance functions are estimated, is its high sensitivity to misspecification of the propensity score model. In a controversial paper \textcite{Kang_Schafer_2007} show in a simulation study that slight misspecification of both nuisance parameters may lead to serious bias in estimating the potential (or missing) outcome. Most importantly this bias can be bigger for AIPW than for estimators that only use the outcome or the treatment model in certain situations. Their paper highlights a major drawback of the estimator for practical empirical research. Even though double robustness is a very desirable theoretical property, it rests on the restrictive assumption of at
least one nuisance parameter to be fully correctly specified. In the context of machine learning estimators dropping below a minimum $o\left(n^{-\frac{1}{4}}\right)$ rate might be counterbalanced by a better convergence rate for the other nuisance parameter if sample splitting is applied (for details see \textcite{Chernozhukov_Chetverikov_Demirer_Duflo_Hansen_Newey_2017}).\\
A second potential problem of AIPW is that by using the inverse prediction of the propensity score directly, very low and very large values of the propensity score lead to extreme weights. Although appropriate from an asymptotic perspective, slight misspecification for such predictions can increase the bias drastically (see also \textcite{Cao_Tsiatis_Davidian_2009}, \textcite{Tan_2010} and \textcite{Vermeulen_Vansteelandt_2015}).\\
General forms of misspecification might be more likely when using very sophisticated prediction methods like machine learning that tend to be sensitive to violations of their major identifying assumptions. In the next two sections we will therefore review alternative suggestions that are less closely related to semiparametric theory but more concerned with the finite sample performance of the estimator.
\subsection{Approximate Residual Balancing}
Approximate Residual Balancing (ARB) proposed by \textcite{Athey_Imbens_Wager_2018} follows from the same estimation structure using the efficient score function. However, instead of weighting the observed outcomes and the outcome model nuisance parameters with the inverse of the propensity score, weights are optimized in order to minimize MSE. The optimal weighting framework most notably goes back to ideas of \textcite{Graham_Campos_Egel_2012}, \textcite{Hainmueller_2012} and \textcite{Zubizarreta_2015}. The intuition behind this direction of research is that a `good' treatment effects estimator should estimate the potential outcomes by reweighting the observed outcomes in a way that the covariate distributions in both treatment samples are identical. Given (\ref{eq:CIA}), if both samples are identical in terms of observable characteristics then calculating the mean difference between the reweighted observed outcomes is a consistent estimator because the samples are then as if they would come from an experiment. However, by balancing the two covariate distributions nearly perfectly the weights across observations may become very volatile. This directly maps into a higher variance of the estimator (\textcite{Zubizarreta_2015}). Thus, the main contribution of this literature is that by controlling the weighting process directly within an optimization that maximizes balancing subject to some second moment upper bound one can achieve good performance in terms of MSE.\\
Consequentely, instead of predicting the propensity score using a machine learner, \textcite{Athey_Imbens_Wager_2018} solely rely on outcome model predictions and then use the structure as in (\ref{eq:aipw}) with bias-variance trade-off optimized weights. For ATE their estimator can be written as
\begin{align}
\hat{\text{ATE}}=&\frac{1}{n}\sum_{i=1}^n\left(\hat{\text{E}}[Y_i|D_i=1,X_i]-\hat{\text{E}}[Y_i|D_i=0,X_i]\right)+\frac{1}{n_1}\sum_{i=1}^nD_i\hat{W}_i^{ARB}(Y_i-\hat{\text{E}}[Y_i|D_i=1,X_i]) \label{eq:ARB} \\
&-\frac{1}{n_0}\sum_{i=1}^n(1-D_i)\hat{W}_i^{ARB}(Y_i-\hat{\text{E}}[Y_i|D_i=0,X_i]). \nonumber
\end{align}
Crucially, the estimator assumes linearity of the outcome regression. This is necessary for two reasons. First of all, the conditional expectations in (\ref{eq:ARB}) can then be estimated using a linear machine learner like the Lasso that guarantees certain convergence rates for a sparse model. Secondly, in order to control the bias that arises from Lasso predictions, one needs the estimator to achieve approximate balancing. This requires that terms like $\Vert\sum_{i=1}^NX_i-\sum_{i=1}^ND_i\hat{W}_i^{ARB}X_i\Vert_{\infty}$ need to be bounded and converge at rates fast enough to counterbalance the decreased convergence rate of the Lasso. By controlling for the weighting scheme directly, such rates can be achieved while still keeping the second moments of the weights from getting too large.\\
Thus, the price to pay for asymptotically optimal MSE weights is the linearity assumption that kicks in twice. Without linearity neither the Lasso rate is sufficient nor is the weighting scheme optimal. We therefore expect ARB to drastically fail in settings where the outcome process is highly non-linear and tree-based methods or neural nets are more appropriate estimators. Strictly speaking the ARB might therefore not even be interpreted as a semiparametric estimator but rather as a flexible linear model with a bias adjustment motivated by the efficient score structure.
\subsection{Augmented Propensity Score Matching}
The previous discussion suggests that there is potential to reweight the observed outcomes in a more sophisticated way than just using the inverse of the estimated propensity score. Essentially we build our considerations on the treatment effects estimation literature where instead of labelling the outcome equation nuisance parameters augmentation terms also takes into account the outcome process as a bias adjustment.\\
Most notably \textcite{Abadie_Imbens_2006} for K-NN matching find that the vanishing bias of their estimator is a function of the outcome model nuisance parameters. Their key finding is that this bias term can converge to zero at a rate potentially slower than $N^\frac{1}{2}$. Based on this observation \textcite{Abadie_Imbens_2011} develop a bias adjusted K-NN matching estimator where the outcome nuisances are estimated using a global nonparametric method and are then equally weighted using the $K$ observations used for estimating the potential outcome. Similar to AIPW estimation their estimator is doubly robust but incorporates the desirable properties of matching. Generally the idea is that instead of basing the reweighting for a single potential outcome prediction on only one particular propensity score prediction, one should use more predictions close by. The distance is defined in terms of the covariates of the other observations. This feature gives matching the very intuitive interpretation of estimating treatment effects using the comparison of similar observations that differ in their treatment status.\\
Another class of estimators incorporates the same principal ideas but instead of using the covariates directly smooths around an estimated propensity score. A first example of this kind of propensity score or Kernel matching was introduced by \textcite{Heckman_Ichimura_Todd_1998}. \textcite{Lechner_Miquel_Wunsch_2011} build on this idea and use a triangular Kernel with data-adaptive bandwidth choice. In reference to \textcite{Dehejia_Wahba_1998} they call their estimator radius matching (RM). Following the findings in \textcite{Abadie_Imbens_2006} they also implement a bias adjustment term that, however, only uses the estimated propensity scores as predictors for the outcome nuisance parameters. Finally \textcite{Abadie_Imbens_2016} derive the asymptotic properties of propensity score matching taking into account the first stage estimation of the propensity score. Unlike the estimator of \textcite{Lechner_Miquel_Wunsch_2011}, their estimator uses a fixed number of matches that are equally weighted. From the perspective of nonparametric estimation the authors establish a result for propensity score matching using a uniform Kernel.\\
The big advantage of all of these estimators is that by smoothing around the best match, they are more robust towards misspecification of a single observation propensity score prediction. Especially for propensity scores close to zero IPW is very prone to large biases since minor misspecification of the treatment effects model lead to large bias weights. Also the critique of \textcite{Kang_Schafer_2007} should only apply to a lesser extent since neither the treatment nor the outcome model have to be completely correctly specified but only the local average around one of the nuisance predictions. Formal results have been shown by \textcite{Heckman_Ichimura_Todd_1998} for parametric and nonparametric estimation of the propensity score. More generally, advances in the field of nonparametric regression with generated data also suggest that the second step Kernel smoothing improves convergence rates (see \textcite{Mammen_Rothe_Schienle_2012}). However, to the extent of the author's knowledge there is no result so far that describes the smoothing properties of a Kernel when faced with a machine learning generated input.\\
Machine learning estimators of the nuisance parameters can be seen as a further advancement in robustifying treatment effects estimators against misspecification, though, none of the estimators considered here are directly suitable for combining with machine learning approaches. Therefore we propose an Augmented Radius Matching (ARM) estimator that like RM uses a triangular Kernel with data-adaptive bandwidth choice to smooth Lasso predicted propensity scores around the best match to reweight the observed outcomes. The same procedure is used to weight the outcome nuisance parameters also predicted by Lasso. Drawing on the structure of (\ref{eq:aipw}) for ATE we now use the weights
\begin{align}\label{eq:ARM}
\hat{W}_i^{ARM}&=K_h\left[\frac{1}{\hat{e}(x)}-\frac{1}{\hat{e}(x_i)}\right]\quad \text{with}\\
K_h(u)&=\max\left(0,1-\frac{\vert u\vert}{h}\right).
\end{align}
To account for outliers the bandwidth $h$ is implicitly chosen such that some multiple of the 90$\%$ quantile and not the maximum of the distribution of the propensity score distance defines a radius determining the maximum distance up to which observations are taken into account. In this sense the weighting strategy represents a sophisticated Kernel matching approach with data-adaptive bandwidth. For some more concrete specifications of the bandwidth choice we refer to the simulation study in \textcite{Huber_Lechner_Wunsch_2013} and just follow the recommended parameter choices further investigated in \textcite{Huber_Lechner_Steinmayr_2015}.\\
We conjecture that by combining the extended double robustness property from the smoothing behaviour with the extended robustness towards functional form misspecification in the first step nuisance predictions ARM performs well in finite sample. Additionally like K-NN matching with bias adjustment it should also exhibit the double robustness property necessary to cope with the decreased convergence rates of machine learning estimators. Unlike ARB the estimator is not fine-tuned to minimize MSE in finite sample which may be a disadvantage. However, by still relying on a propensity score weighting, it does not have to make specific assumptions about the linearity of the outcome model and therefore in principle would allow to use non-linear machine learning estimators like random forests or neural nets. Seen like this our estimator represents a compromise between the neat structure of AIPW and weighting concerns that are likely to become more prominent when highly complex methods as machine learning techniques are used.
\subsection{Refitted Doubly Selected Propensity Score Estimators}
The variable selection property of Lasso enables to construct an additional class of estimators that builds upon the double selection suggestions in \textcite{Belloni_Chernozhukov_Hansen_2014}. Instead of using the union of selected covariates from Lasso regression on outcome and treatment to select the relevant covariates for the outcome model, one may use this set to refit the propensity score model (see also \textcite{Athey_Imbens_Wager_2018}).
\begin{enumerate}
\item Run a Lasso regressions on $Y$ and $D$ and save the covariate sets $X^Y$ and $X^D$ selected.
\item Construct the union $X^{sel}=X^Y\cup X^D$.
\item Predict the propensity score using for example a logit model with $P(D=1|X^{sel}=x)$.
\item Use the propensity score predictions to form IPW or RM to estimate potential outcomes and treatment effects.
\end{enumerate}
The validity of this approach follows the same reasoning as the double selection procedure for the outcome equation. Since for consistency it does not matter if one controls for all relevant confounders in the propensity score or the outcome model, the estimator should be able to cope with the concerns of single selection discussed in section \ref{sec:naive}. However, it is unlikely that the estimator achieves semiparametric efficiency because the propensity score is not estimated nonparametrically. In particular, by including the covariates from the outcome model the refitted parametric model does not approximate a global nonparametric method. An additional disadvantage of this estimator is that it does not allow to use any other machine learning estimator than the Lasso since it explicitly depends on the variable selection feature. Though, from an economic perspective this can also be interpreted as a strength of the estimator because it allows the researcher to gain some intuition what channels should be controlled for when trying to identify causal effects.
\section{A Monte Carlo Experiment}\label{sec:montecarlo}
\subsection{Setup and DGPs}
To assess the finite sample behaviour of the estimators discussed, we use a specific Monte Carlo set-up. In general the data-generating process (DGP) employed is loosely inspired by designs in \textcite{Busso_DiNardo_McCrary_2014} for low-dimensional and \textcite{Athey_Imbens_Wager_2018} and \textcite{Belloni_Chernozhukov_Hansen_2014} for high-dimensional settings. In particular, we model the selection into treatment with an index model $D_i=I(D_i^*>0)$ where $D_i^*$ follows $D_i^*=\beta_d X_i+c_dv$ where $i=1,...,n$ with $n=2000$ and the dimension of the covariate space is $p=2000$. Similarly the outcome is modelled using $Y_i=D_i\theta+D_iX_i\beta_{g1}+(1-D_i)X_i\beta_{g0}+c_yu_i$.\\
We set the error terms as $\begin{pmatrix}u\\v\end{pmatrix}\sim N(0,I)$ and determine the feature matrix as jointly normal such that $X\sim N(0,\Sigma)$. For most specifications a Toeplitz structure of the variance-covariance matrix is assumed. In particular every element of $\Sigma$ obeys $\sigma_{jk}=q^{\left|j-k\right|}$ where $j$ and $k$ are row and column indices. Notice that the ATE can be explicitly derived from the model as
\begin{align*}
\text{ATE}=&E[Y^1-Y^0]
=E[\theta+(\beta_{g1}-\beta_{g0})X]
=\theta
\end{align*}
since the first moment of $X$ is zero. However, this does not hold for the ATET except for the case when $\beta_{g1}=\beta_{g0}$ which we exclude such that we obtain effect heterogeneity.\\
Finally, $c_d$ and $c_y$ are scaling factors that are chosen in order to achieve a certain $R^2$ for the two processes. While there is a closed form solution for the treatment process, the parameter value for the outcome process has to be simulated.\\
With the Monte Carlo study we want to approximate four fundamental questions concerning the finite sample performance of the estimators. 
\begin{itemize}
\item What is the effect of the choice of the Lasso tuning parameter on the performance of the estimators?
\item How do the different estimators perform under deteriorating qualities of nuisance parameter prediction?
\item In particular, how do the estimators perform when the covariates exhibit certain correlation structures?
\item How do the estimators cope with unknown experimental settings?
\end{itemize}
Most prominently the relative within design performance of the different estimators shall be compared. Table \ref{tab:mcdesigns} depicts the different specifications used.\\
\begin{table}[h!]
\centering
\caption{Monte Carlo designs overview}
\label{tab:mcdesigns}
\begin{threeparttable}
\begin{tabular}{l|cccc}
Design & \multicolumn{2}{c}{sparsity} & correlation structure & parallel \\
& outcome & treatment & & \\ \midrule
1 & sparse & sparse & Toeplitz & yes \\
2 & moderate & moderate & Toeplitz & yes \\
3 & dense & dense & Toeplitz & yes \\
4 & sparse & moderate & Toeplitz & yes \\
5 & moderate & sparse & Toeplitz & yes \\
6 & sparse & dense & Toeplitz & yes \\
7 & dense & sparse & Toeplitz & yes \\
8 & sparse & sparse & randomized clusters & yes \\
9 & sparse & sparse & ordered clusters & yes \\
10 & sparse & sparse & Toeplitz & no \\
\bottomrule \bottomrule
\end{tabular}
\begin{tablenotes}
\small
\item Monte Carlo designs with 1000 repetitions for the case $n=2000$ and $p=2000$. Model variance parameters are chosen such that $R^2_Y\in [0.2,0.8]$ and $R^2_D\in [0,0.5]$.
\end{tablenotes}
\end{threeparttable}
\end{table}
\subsection{Results}
\subsubsection{Choice of the Tuning Parameter}
\begin{figure}[h!]
\caption{RMSE and tuning parameter choice for different estimators in design 1}
\label{fig:tune_design1}
\begin{subfigure}{.5\textwidth}
\includegraphics[scale=0.5]{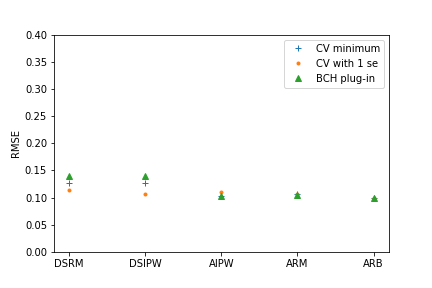}
\caption{$R^2_D=0.5$, $R^2_Y=0.2$}
\end{subfigure}
\begin{subfigure}{.5\textwidth}
\includegraphics[scale=0.5]{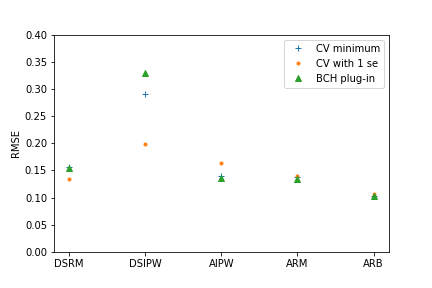}
\caption{$R^2_D=0.5$, $R^2_Y=0.8$}
\end{subfigure}
\end{figure}

Before considering the performance of the different estimators, one should clarify the question under which machine learning estimator the comparison is made. In fact there are as many Lasso estimators as there are tuning parameter choices. Therefore the three different bandwidth choices discussed in section \ref{sec:naive} are compared in the ideal setting of design 1. The treatment and the outcome process are modelled with $\beta_{j,d}=\beta_{j,g0}=\left(\frac{1}{j}\right)^2$ and $\beta_{j,g1}=\left(\frac{1}{s_yj}\right)^2$ where the index $j$ denotes the column of the covariate in the feature matrix and $s_y\neq 1$ is an arbitrarily chosen parameter to make the effects heterogeneous. Under this DGP the prediction problem is approximately sparse and we expect Lasso based methods to work well. Indeed this is what figure \ref{fig:tune_design1} depicts. Under their best bandwidth choice all methods show good performance. While for the double selection (DSRM, DSIPW) approaches the 1-standard-error rule seems to be dominant, the nuisance prediction approaches (AIPW, ARM, ARB) seem to work best with either the theoretically justified or the cross-validation choice. Intuitively, by fitting a smaller model the 1-standard-error rule Lasso trades a bit more bias against hopefully less variance out of sample. Since in principle the union of both covariates sets would not be necessary to achieve consistency, variable selection with the union is prone to including too much variables in the model and a higher penalty is optimal. Thus, by explicitly relying on the model selection instead of the prediction feature, cross-validation minimization cannot be optimal if it is optimal for prediction. For all further specifications it turns out that the pattern prevails with the exception that the theoretically justified tuning parameter leads to more and more arbitrary results as the Lasso assumptions get violated. Since the plug-in tuning parameter in no setting strongly dominates the cross-validation based methods, we proceed by comparing the performance of DSRM and DSIPW using the 1-standard-error rule and AIPW, ARM and ARB using the minimized cross-validation criterion.
\subsubsection{Finite Sample Behaviour Under Ideal Specification}
\begin{figure}[h!]
\caption{Finite sample distribution of double selected and double machine learning estimators}
\label{fig:hist_design1}
\begin{subfigure}{.5\textwidth}
\includegraphics[scale=0.5]{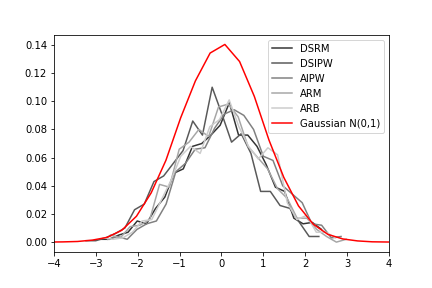}
\caption{$R^2_D=0.5$, $R^2_Y=0.2$}
\end{subfigure}
\begin{subfigure}{.5\textwidth}
\includegraphics[scale=0.5]{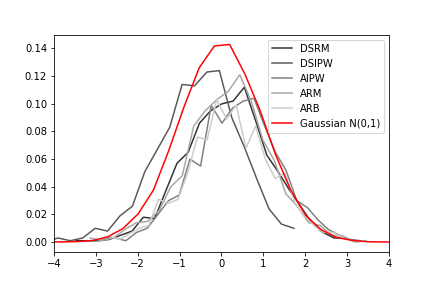}
\caption{$R^2_D=0.5$, $R^2_Y=0.8$}
\end{subfigure}
\end{figure}

Given these a priori considerations, table \ref{tab:mcresults1} shows that methods which take into account both the treatment and the outcome process perform comparatively well. As already shown graphically in the motivation part and in line with theoretical claims, methods based on simply predicting the propensity score (we also use cross-validation minimization) and using this as the only nuisance are heavily biased. A second observation is that alternative weighting schemes in the context of doubly robust estimators perform better than IPW based methods. We suspect the enhanced smoothing properties of the estimators to be more suitable for the lower machine learning convergence rates. Among the estimators based on the efficient score structure ARB exhibits the best performance in terms of RMSE and ARM has the lowest bias.\\
Figure \ref{fig:hist_design1} shows that the standardized draws from the estimators studied are pretty close to 1000 draws from the standard normal distribution. This indicates that despite potential differences in asymptotic efficiency all estimators are nicely behaved and it is likely that they converge towards a normal distribution.
\subsubsection{Misspecification}
\begin{figure}[t!]
\caption{Augmentation term mechanisms for ARM and AIPW in design 2}
\label{fig:aug_design2}
\begin{subfigure}{.5\textwidth}
\includegraphics[scale=0.5]{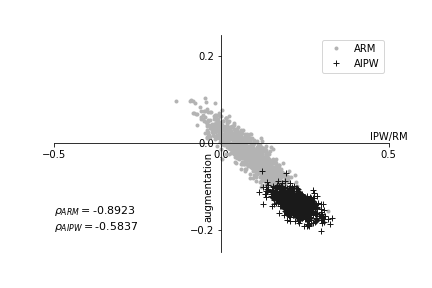}
\caption{$R^2_D=0.5$, $R^2_Y=0.2$}
\end{subfigure}
\begin{subfigure}{.5\textwidth}
\includegraphics[scale=0.5]{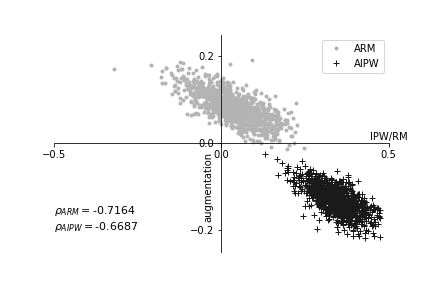}
\caption{$R^2_D=0.5$, $R^2_Y=0.8$}
\end{subfigure}
\vspace{128in}
\end{figure}

\begin{sidewaystable}[ph!]
\centering
\small
\caption{Monte Carlo results}
\label{tab:mcresults1}
\begin{threeparttable}
\begin{tabular}{llllllllllllllll}
 & & \multicolumn{7}{c}{$R^2_Y=0.2$} & \multicolumn{7}{c}{$R^2_Y=0.8$}\\
\cmidrule(lr){3-9} \cmidrule(lr){10-16}
Design & &      RM  &      IPW  &      DSRM  &      DSIPW  &      AIPW  &      ARM  &      ARB  &      RM  &      IPW  &      DSRM &      DSIPW &      AIPW &      ARM &      ARB \\
\midrule
1  &  bias &  0.1811 &  0.3979 &  0.0059 &  0.0243 &  0.0223 &  0.0008 &  0.0104 &  0.0673 &  0.5228 &  0.0184 &  0.0963 &  0.0465 &  0.0205 &  0.0214 \\
  &    se &  0.2616 &  0.1229 &  0.1145 &  0.1037 &  0.1006 &  0.1069 &  0.0985 &  0.1927 &  0.1516 &  0.1338 &   0.173 &  0.1324 &  0.1371 &  0.1012 \\
  &   \#NA &     0 &     0 &     0 &     0 &     0 &     0 &     0 &     0 &     0 &     0 &     0 &     0 &     0 &     0 \\
  &  RMSE &  0.3182 &  0.4165 &  0.1146 &  0.1065 &  0.1031 &  0.1069 &  0.0991 &  0.2041 &  0.5444 &   0.135 &   0.198 &  0.1403 &  0.1387 &  0.1034 \\
2  &  bias &  0.0782 &  0.2205 &  0.0649 &  0.0892 &   0.082 &  0.0575 &  0.0433 &  0.0368 &  0.3433 &  0.0381 &  0.1353 &  0.2137 &  0.1193 &  0.1726 \\
  &    se &  0.0648 &  0.0339 &  0.0373 &  0.0405 &  0.0276 &  0.0318 &   0.028 &  0.0765 &  0.0542 &  0.0745 &  0.2066 &  0.0407 &  0.0569 &  0.0361 \\
  &   \#NA &     0 &     0 &     0 &     0 &     0 &     0 &     0 &     0 &     0 &     1 &     1 &     0 &     0 &     0 \\
  &  RMSE &  0.1015 &  0.2231 &  0.0748 &   0.098 &  0.0865 &  0.0657 &  0.0516 &  0.0849 &  0.3475 &  0.0837 &   0.247 &  0.2175 &  0.1322 &  0.1763 \\
3  &  bias &  1.3119 &  2.6586 &  0.0627 &  0.4772 &  1.4263 &   1.158 &  0.8598 &  0.6393 &  3.5018 &  1.4383 &  6.2575 &  3.0592 &  1.7353 &   3.241 \\
  &    se &  0.6647 &  0.3404 &  0.3785 &  0.5662 &  0.2885 &  0.3303 &    0.27 &  1.0282 &  0.7218 &  1.2293 &  5.6904 &  0.5285 &  0.8082 &  0.3269 \\
 &   \#NA &     0 &     0 &     0 &     0 &     0 &     0 &     0 &     0 &     0 &   997 &   997 &     0 &     0 &     0 \\
 &  RMSE &  1.4707 &  2.6803 &  0.3837 &  0.7405 &  1.4552 &  1.2042 &  0.9012 &  1.2107 &  3.5754 &  1.8921 &  8.4579 &  3.1045 &  1.9143 &  3.2574 \\
4 &  bias &   0.474 &  0.4149 &  0.0116 &  0.0714 &  0.1701 &  0.1551 &  0.1105 &  0.4944 &   0.412 &  0.0138 &   0.092 &  0.0788 &   0.057 &  0.0536 \\
 &    se &  0.6811 &  0.3964 &  0.6541 &  0.7548 &  0.3885 &  0.6251 &  0.4063 &  0.3163 &  0.1776 &  0.2774 &  0.3327 &  0.1564 &  0.1975 &   0.127 \\
 &   \#NA &     0 &     0 &     0 &     0 &     0 &     0 &     0 &     0 &     0 &     0 &     0 &     0 &     0 &     0 \\
 &  RMSE &  0.8298 &  0.5739 &  0.6542 &  0.7581 &  0.4241 &   0.644 &  0.4211 &  0.5869 &  0.4486 &  0.2777 &  0.3452 &  0.1751 &  0.2055 &  0.1379 \\
5 &  bias &  0.0521 &  0.0945 &  0.0117 &  0.0256 &  0.0641 &  0.0375 &  0.0328 &  0.0499 &  0.0933 &  0.0938 &  0.1258 &  0.0225 &  0.0177 &  0.0158 \\
 &    se &  0.1145 &  0.0877 &  0.1118 &  0.1102 &  0.0859 &  0.1076 &  0.0876 &  0.0496 &  0.0368 &  0.0506 &  0.0686 &  0.0247 &  0.0284 &  0.0244 \\
 &   \#NA &     0 &     0 &     0 &     0 &     0 &     0 &     0 &     0 &     0 &     0 &     0 &     0 &     0 &     0 \\
 &  RMSE &  0.1258 &   0.129 &  0.1124 &  0.1132 &  0.1072 &  0.1139 &  0.0935 &  0.0703 &  0.1003 &  0.1065 &  0.1433 &  0.0334 &  0.0334 &  0.0291 \\
6 &  bias &  0.6953 &  0.3377 &  0.0032 &  0.0717 &  0.1022 &  0.1449 &  0.0668 &  1.7947 &  0.7162 &  0.0371 &  1.5609 &   0.181 &  0.4649 &  0.1167 \\
 &    se &  0.3998 &  0.2524 &  0.3559 &  0.3495 &  0.2462 &  0.3268 &  0.2581 &   1.482 &  0.6665 &  1.3181 &  3.5309 &  0.4999 &  1.2243 &  0.2974 \\
 &   \#NA &     0 &     0 &     0 &     0 &     0 &     0 &     0 &     0 &     0 &   993 &   993 &     0 &     0 &     0 \\
 &  RMSE &   0.802 &  0.4216 &   0.356 &  0.3567 &  0.2666 &  0.3575 &  0.2666 &  2.3275 &  0.9784 &  1.3186 &  3.8605 &  0.5316 &  1.3096 &  0.3195 \\
7 &  bias &  0.1031 &  0.3188 &     nan &     nan &  0.0182 &  0.0157 &  0.0146 &  0.1101 &  0.5164 &     nan &     nan &  0.0782 &  0.0733 &  0.0702 \\
 &    se &  0.3373 &  0.2466 &     nan &     nan &  0.1209 &  0.1206 &  0.1156 &  0.3086 &    0.24 &     nan &     nan &  0.1272 &   0.128 &  0.1221 \\
 &   \#NA &     5 &     0 &  1000 &  1000 &     0 &     5 &     0 &     0 &     0 &  1000 &  1000 &     0 &     0 &     0 \\
 &  RMSE &  0.3528 &   0.403 &     nan &     nan &  0.1223 &  0.1216 &  0.1166 &  0.3276 &  0.5695 &     nan &     nan &  0.1493 &  0.1475 &  0.1408 \\
8 &  bias &  0.0362 &  0.4543 &  0.1355 &  0.0398 &  0.2071 &  0.0749 &  0.0967 &  0.0512 &  0.4463 &  0.0401 &  0.1401 &  0.0717 &  0.0407 &  0.0348 \\
 &    se &  0.5688 &  0.4814 &  0.5526 &  0.5338 &  0.4857 &  0.5553 &  0.4824 &  0.2363 &  0.2257 &  0.2091 &  0.2464 &  0.1998 &  0.2106 &  0.1577 \\
 &   \#NA &     0 &     0 &     0 &     0 &     0 &     0 &     0 &     0 &     0 &     0 &     0 &     0 &     0 &     0 \\
 &  RMSE &  0.5699 &  0.6619 &   0.569 &  0.5352 &   0.528 &  0.5603 &   0.492 &  0.2418 &  0.5001 &   0.213 &  0.2834 &  0.2123 &  0.2145 &  0.1615 \\
9 &  bias &  0.1291 &  0.5224 &  0.0048 &  0.3057 &  0.1679 &  0.0233 &  0.0786 &  0.1194 &  0.5251 &   0.006 &  0.3069 &  0.0444 &  0.0051 &  0.0279 \\
 &    se &  0.8584 &  0.7466 &  0.8386 &  0.9142 &  0.7364 &  0.8466 &  0.7161 &  0.3538 &  0.3821 &  0.3282 &  0.4626 &  0.3237 &  0.3359 &   0.223 \\
 &   \#NA &     0 &     0 &     0 &     0 &     0 &     0 &     0 &     0 &     0 &     0 &     0 &     0 &     0 &     0 \\
 &  RMSE &   0.868 &  0.9112 &  0.8386 &   0.964 &  0.7553 &  0.8469 &  0.7204 &  0.3734 &  0.6494 &  0.3283 &  0.5551 &  0.3267 &   0.336 &  0.2247 \\
10 &  bias &  0.0027 &  0.0022 &  0.0001 &  0.0014 &  0.0024 &  0.0013 &  0.0012 &  0.0003 &  0.0025 &  0.0004 &  0.0026 &  0.0022 &   0.001 &  0.0004 \\
 &    se &   0.185 &  0.1613 &  0.1299 &  0.0973 &  0.0951 &  0.1007 &  0.0984 &  0.2247 &  0.1793 &  0.1634 &   0.132 &  0.0998 &  0.1079 &  0.1028 \\
 &   \#NA &     0 &     0 &     0 &     0 &     0 &     0 &     0 &     0 &     0 &     0 &     0 &     0 &     0 &     0 \\
 &  RMSE &   0.185 &  0.1613 &  0.1299 &  0.0973 &  0.0951 &  0.1007 &  0.0984 &  0.2247 &  0.1793 &  0.1634 &  0.1321 &  0.0999 &  0.1079 &  0.1028 \\
\bottomrule \bottomrule
\end{tabular}
\end{threeparttable}
\begin{tablenotes}
\footnotesize
\item Monte Carlo designs with 1000 repetitions for the case $n=2000$ and $p=2000$. Model variance parameters are chosen such that $R^2_Y\in [0.2,0.8]$ and $R^2_D=0.5$. The different estimators considered are radius matching (RM), inverse probability weighting (IPW), double selected radius matching and inverse probability weighting (DSRM,DSIPW), augmented inverse probability weighting and radius matching (AIPW,ARM) and approximate residual balancing (ARB).
\end{tablenotes}
\end{sidewaystable}
In designs 2-9 we consider two different forms of misspecification. Instead of including unobserved effects in the model, we introduce misspecification as coming from the deteriorating quality of nuisance parameter prediction. For the Lasso this may be achieved by violating the sparsity (designs 2-7) or the restricted eigenvalue assumption (designs 8 and 9). Before examining the latter by analysing the effects of clusters in the feature variance-covariance matrix, the processes are specified more `densely'.\\
For both the treatment and the outcome process a moderately dense process is characterized by coefficients of the form $\beta_j=\left(\frac{1}{j+10}\right)$ and a dense process by $\beta_j=\sqrt{\frac{1}{j}}$. We notice that the dense case represents an upper bound for the degree of misspecification since for such a DGP the asymptotic variance of the estimator becomes unbounded. Moving from the neat design 1 to these misspecified cases for both processes simultaneously shows that the performance of the estimator first decreases slowly before RMSEs explode. For the moderately misspecified design 2 a similar pattern as in design 1 emerges such that IPW-based estimators are again dominated by the alternative double type estimators. Design 2 represents a more realistic situation where not all assumptions necessary to achieve good predictions are specified but estimators are only moderately biased. In such a setting ARM and ARB perform particularly well and dominate AIPW. The enhanced performance of ARM stems from two sources. Most importantly the smoothing property of the estimator allows a higher degree of misspecification because predicted counterfactual outcomes are based on more than one propensity score prediction. Indeed figure \ref{fig:aug_design2} shows that RM not using any adjustment term is probably inconsistent but performs way better than IPW. Furthermore this property also improves the double robustness mechanism of the estimator. Under both DGPs the correlations between the augmentation term and the propensity score based estimator are stronger for ARM. Thus, whenever RM exhibits positive bias the augmentation term corrects for this by adding a negative term and vice versa. Again the smoothed propensity score allows to give higher weights to the outcome nuisances in areas of the support where it is actually needed most.\\
As the results for design 4 and 5 indicate a potential problem with ARM is its sensitivity regarding the choice of the bandwidth. While it beats AIPW in terms of bias, it is worse in terms of RMSE. A higher bandwidth that trades a bit of bias against a decrease in variance might therefore more appropriate in such settings. Moreover, double selection based estimators exhibit an increasing number of draws where they cannot be estimated at all. Since choosing between the different covariates becomes increasingly difficult for the Lasso, the probability that the union of the two sets consists of very many variables becomes bigger such that the refitted propensity score model can only hardly be estimated as in designs 3,6 and 7.\\
Having either of the two processes to be strongly misspecified (designs 6 and 7), forces the estimator to heavily rely on only one of the processes. When the propensity score is very hard to predict, the matching based models become increasingly worse. Reweighting the outcomes and the predicted outcome nuisances with weights that depend on smoothed distances between propensity scores makes these weights increasingly useless. ARB is particularly well-suited for design 6 since it does not depend on the treatment model and shows decent performance. For design 7 AIPW, ARM and ARB perform similarly. This is in line with theoretical considerations because AIPW and ARM mostly depend on the propensity score weighting and ARB should not exhibit any relative losses as long as the outcome model is linear.\\
Besides the violation of the sparsity assumption, nuisance prediction may also suffer from correlation clusters in the feature variance-covariance matrix. This case may be even more relevant in practice since often high dimensions in covariate spaces are generated by either including classes of variables that actually capture the same economic channels or by just generating interactions and polynomials out of an existing dataset. While especially in the latter case one may believe in the sparsity of the model, variables will exhibit clustered correlation structures by construction. Simulations designs 8 and 9 were generated with stochastic correlation matrices using the method by \textcite{Hardin_Garcia_Golan_2013}. Since design 9 puts the cluster structure on neighbouring covariates with similar importance for outcome and treatment, not much additional bias compared to design 1 is expected. However, the standard errors increase because the nuisance prediction correlation between Monte Carlo draws increases.\\
Design 8 puts the cluster structure randomly such that potentially very different covariates in terms of effects become strongly correlated reflecting the dimensionality scaling discussed before. In general a larger bias compared to design 9 is observed since Lasso now becomes partly unable to discriminate covariates with different effects while standard errors increase less drastic. AIPW, ARM and ARB perform similarly but dominate the double selection based estimators. Again ARM and ARB are significantly less biased than AIPW. However, the differences are not as drastic as in the sparsity based misspecification.
\subsubsection{Unknown Experiments}
\begin{center}
\begin{table}[h!]
\centering
\caption{Monte Carlo results for experimental design 1}
\label{tab:mcresults_exper}
\centering
\begin{threeparttable}
\begin{tabular}{lllllllll}
& \multicolumn{8}{c}{$R^2_Y=0.2$} \\
\cmidrule(lr){2-9}
& Naive & RM  &      IPW  &      DSRM  &      DSIPW  &      AIPW  &      ARM  &      ARB \\
\midrule
bias & 0.0017 & 0.0064 &   0.551 &  0.0857 &  0.1785 &  0.5458 &  0.0003 &  0.3525 \\
se & 0.3524 & 5.1437 &  4.4212 &  5.0931 &  4.4788 &  4.4189 &  5.1392 &  4.9091 \\
\#NA & 0 &    0 &     0 &    44 &    44 &     0 &     0 &     0 \\
RMSE & 0.3524 & 5.1437 &  4.4554 &  5.0938 &  4.4823 &  4.4525 &  5.1392 &  4.9217\\
\bottomrule
\end{tabular}
\begin{tabular}{lllllllll}
& \multicolumn{8}{c}{$R^2_Y=0.8$}\\
\cmidrule(lr){2-9}
& Naive &    RM  &      IPW  &      DSRM &      DSIPW &      AIPW &      ARM &      ARB \\
\midrule
bias & 0.0021 & 0.0344 &  0.5119 &   0.0444 &   0.0241 &   0.512 &   0.0209 &  0.2873 \\
se & 0.1706 & 5.9639 &  4.8439 &   5.9327 &   5.2707 &  4.8462 &   5.9495 &  5.2581 \\
\#NA & 0 &   0 &     0 &      0 &      0 &     0 &      0 &     0 \\
RMSE & 0.1706 & 5.964 &  4.8708 &   5.9328 &   5.2708 &  4.8732 &   5.9495 &   5.266 \\
\bottomrule\bottomrule
\end{tabular}
\begin{tablenotes}
\footnotesize
\item Monte Carlo designs with 1000 repetitions for the case $n=2000$ and $p=2000$. Model variance parameters are chosen such that $R^2_Y\in [0.2,0.8]$ and $R^2_D=0$. The different estimators considered are radius matching (RM), inverse probability weighting (IPW), double selected radius matching and inverse probability weighting (DSRM,DSIPW), augmented inverse probability weighting and radius matching (AIPW,ARM), approximate residual balancing (ARB) and the naive estimator $\hat{\text{ATE}}=\frac{1}{n_1}\sum_{i=1}^ND_iY_i-\frac{1}{n_0}\sum_{i=1}^N(1-D_i)Y_i$.
\end{tablenotes}
\end{threeparttable}
\end{table}
\end{center}

We now consider the interesting DGP that assigns treatment completely at random. Hence, we model the case of an experiment in which the researcher is actually unaware of the fact that treatment is assigned randomly and therefore estimates the treatment effect using one of the discussed selection on observables methods. In particular we model $D^*=v$ such that given this knowledge no covariate is necessary to consistently estimate the treatment effect. In such a setting all estimators with the exception of the radius matching based estimators exhibit considerable bias. The reason is that by smoothing over regions with comparable propensity scores, radius matching should weight the different observations nearly equally, while IPW based methods where the weight one observation receives only depends on the propensity score prediction for this particular observation fail to equalize weights over the sample. Therefore for settings where $R^2_D$ is close to zero, smoothing methods have a built-in mechanism that enables them to be near the simple unconditional mean difference estimator which would be the appropriate estimator in such settings. The double selection estimators in general also exhibit comparatively low bias as they include unnecessary variables that however do not hurt when estimating ATE. However, unsurprisingly all selection on observables estimators fail in comparison to the simple difference in means estimator.\\
The discussion highlights a major problem of plugging in machine learning predictions in the efficient score or its radius matching based differences as not only unnecessarily many covariates enter the estimator but they heavily distort the reweighting scheme.
\section{Conclusion and Further Research}
In this paper we analysed and reviewed the doubly robust treatment effects estimator structure for average treatment effects estimation. In the context of machine learning this particular estimation structure turns out to be suitable for incorporating machine learning by combining less then $\sqrt{n}$-convergence rates for both the treatment and the outcome model. However, instead of relying on IPW weights that follow from purely asymptotic arguments we have argued for more sophisticated weighting schemes that are more robust to misspecification -- a problem potentially very prevalent for sophisticated machine learning estimators. Within our simulation design it turned out that a weighting scheme that is based on Kernel matching performs well in finite sample. In contrast to the other AIPW alternatives considered the proposed estimator should also be feasible for nonlinear specifications of the nuisance parameter by still relying on propensity score estimation.\\
For the sake of brevity, many aspects of machine learning approaches for treatment effects estimation were not considered in this study. While the goal of our Monte Carlo designs was to examine some theoretically implied properties, it would be particularly interesting to investigate the performance of the different weighting schemes in less artificial settings. As a further practical concern the weak stability of model selection based estimators highlights the need for variable importance measures for other supervised machine learning estimators. Although from a statistical point of view only the predictive performance for the nuisance function estimators matters, economists are nevertheless interested in developing a feeling for the different driving forces behind the models.\\
Further any rigorous asymptotic analysis of Kernel weighting schemes would require the exact convergence rates of nonparametric methods with machine learning generated inputs. Additional research in this field seems to be very promising.
\newpage
\printbibliography

\newpage
\appendix
\section{Consistency of AIPW and IPW} \label{app:aipw}
Similar consistency results for the AIPW etimator are for example given in \textcite[chapter 13]{Tsiatis_2006}. We depict the analysis here for better readability.\\
From considerations in section \ref{sec:doublyrobust} it follows that the ATE for which (\ref{eq:aipw}) is a natural estimator can be written as
\begin{align*}
\text{ATE}=&E\Bigg[\frac{DY}{e(X)}-\frac{(1-D)Y}{1-e(X)}+E[Y|D=1,X]-E[Y|D=0,X]-\frac{DE[Y|D=1,X]}{e(X)}\\
\\
&+\frac{(1-D)E[Y|D=0,X]}{1-e(X)}\Bigg]
\\
=&E\Bigg[E\Bigg(\frac{DY}{e(X)}-\frac{(1-D)Y}{1-e(X)}+E[Y|D=1,X]-E[Y|D=0,X]-\frac{DE[Y|D=1,X]}{e(X)}\\
\\
&+\frac{(1-D)E[Y|D=0,X]}{1-e(X)}\Bigg|X\Bigg)\Bigg]\quad \text{by the Law of Iterated Expectations}
\\
=&E\Bigg[\frac{E[DY|X]}{e(X)}-\frac{E[(1-D)Y|X]}{1-e(X)}+E[Y|D=1,X]-E[Y|D=0,X]\\
\\
&-\frac{E[D|X]E[Y|D=1,X]}{e(X)}+\frac{E[(1-D)|X]E[Y|D=0,X]}{1-e(X)}\Bigg]
\end{align*}
which follows since the nuisance parameter $\eta=(e(X),E[Y|D=1,X],E[Y|D=0,X])$ consists of functions of $X$. Using the Stable Unit Treatment Value Assumption (SUTVA) $Y=DY(1)+(1-D)Y(0)$ yields
\begin{align*}
\text{ATE}=&E\Bigg[\frac{E[DDY(1)+D(1-D)Y(0)|X]}{e(X)}-\frac{E[(1-D)DY(1)+(1-D)(1-D)Y(0)|X]}{1-e(X)}\\
\\
&+E[Y|D=1,X]-E[Y|D=0,X]-\frac{E[D|X]E[Y|D=1,X]}{e(X)}+\frac{E[(1-D)|X]E[Y|D=0,X]}{1-e(X)}\Bigg]\\
\\
=&E\Bigg[\frac{E[D|X]E[Y(1)|X]}{e(X)}-\frac{E[(1-D)|X]E[Y(0)|X]}{1-e(X)}+E[Y|D=1,X]-E[Y|D=0,X]\\
\\
&-\frac{E[D|X]E[Y|D=1,X]}{e(X)}+\frac{E[(1-D)|X]E[Y|D=0,X]}{1-e(X)}\Bigg].
\end{align*}
Then rearranging the first term as
\begin{align*}
\frac{E[D|X]E[Y(1)|X]}{e(X)}=&\frac{E[D|X]E[Y(1)|X]+e(X)E[Y(1)|X]-e(X)E[Y(1)|X]}{e(X)}\\
\\
=&E[Y(1)|X]+\frac{\left(E[D|X]-e(X)\right)E[Y(1)|X]}{e(X)}
\end{align*}
and similarly the second term, leads to
\begin{align*}
\text{ATE}=&E\Bigg[\underbrace{E[Y(1)|X]-E[Y(0)|X]}_{i}+\underbrace{\frac{\left(E[D|X]-e(X)\right)\left(E[Y(1)|X]-E[Y|D=1,X]\right)}{e(X)}}_{iia}\\
\\
&+\underbrace{\frac{\left(E[D|X]-e(X)\right)\left(E[Y(0)|X]-E[Y|D=0,X]\right)}{1-e(X)}}_{iib}\Bigg].
\end{align*}
It follows that AIPW identifies ATE if $iia$ and $iib$ are zero in expectation. From the structure of the terms this is the case if either the propensity score or the two outcome equation estimators are consistent estimators.\\
\\
Using the same arguments for the analysis of IPW reveals that
\begin{align*}
\text{ATE}=&E\Bigg[\frac{DY}{e(X)}-\frac{(1-D)Y}{1-e(X)}\Bigg]\\
\\
=&E\Bigg[\underbrace{E[Y(1)|X]-E[Y(0)|X]}_{i}+\underbrace{\frac{(E[D|X]-e(X))E[Y(1)|X]}{e(X)}}_{iia}+\underbrace{\frac{(E[D|X]-e(X))E[Y(0)|X]}{1-e(X)}}_{iib}\Bigg].
\end{align*}
Again the estimator identifies ATE if $iia$ and $iib$ are zero in expectation. This requires that the propensity score model is correctly specified.
\section{Specification Details for the Monte Carlo Experiments}
All estimators require the Lasso estimation of a treatment and one or two outcome models. The outcome model is always implemented using a linear functional form (penalized OLS), while the binary treatment process is modelled using the logit representation of the Lasso.\\
DSIPW and DSRM are implemented using the union of the selected Lasso features coming from outcome and treatment models. The tuning parameter for these estimators is chosen using the 1-standard-error rule. The propensity score is then estimated with a standard logit regression using the union of all features that were previously selected by the two Lasso regressions. All nuisance parameters are estimated with the \textsc{glmnet} and \textsc{hdm} packages in \textsc{R}. For the Lasso estimation with \textsc{glmnet} we use 10-fold cross-validation and for the Lasso estimation with the plug-in tuning parameter of \textcite{Belloni_Chernozhukov_Hansen_2014} we follow the recommended settings of the \textsc{hdm} package.\\
IPW and AIPW are implemented using their normalized form such that weights sum up to one. Here the outcome Lasso regression is estimated separately for the treated and the untreated sample.\\
Concerning the nuisance parameters RM and ARM are implemented like IPW and AIPW. However, the estimator requires some more detailed choices. Most notably one has to choose up to which range other observations are taken into account to predict the potential outcomes of an observation $i$. Following the implementation as in \textcite{Huber_Lechner_Wunsch_2013} we use the $90\%$ quantile of the of the propensity score distance distribution for observation $i$ and multiply it with a certain factor. Rather arbitrarily but within the recommended range of \textcite{Huber_Lechner_Steinmayr_2015} we specify this factor as $1.5$.\\
ARB is estimated using the \textsc{R} implementation of Stefan Wager. We follow the recommendations as in \textcite{Athey_Imbens_Wager_2018} with the exception that we only consider the Lasso and allow for the tuning parameter choice of \textcite{Belloni_Chernozhukov_Hansen_2014}.
\newpage
\section*{Figures}
\end{document}